\documentclass{aa}

\usepackage{natbib}
\usepackage{graphicx}
\usepackage{times}
\usepackage{amsmath}
\usepackage{txfonts}
\newcommand{\eg} {{\em e.g.}}
\newcommand{\unzero}{$1_\mathrm{0,1}-0_{0,0}$}
\newcommand{\unun}{$1_{1,1}-1_{1,0}$}

\newcommand{\msun}{{\,\rm M}_{\odot}}

\newcommand{\kms}{\,{\rm km\,s}^{-1}}

\newcommand{\nht}{\ifmmode {{\rm NH}_3} \else {NH{\bas 3}} \fi}
\newcommand{\tco}{\ifmmode {^{13}{\rm CO}} \else {$^{13}{\rm CO}$}\fi}
\newcommand{\dco}{\ifmmode {^{12}{\rm CO}} \else {$^{12}{\rm CO}$}\fi}
\newcommand{\cdo}{\ifmmode {{\rm C}^{18}{\rm O}} \else {${\rm C}^{18}{\rm O}$}\fi}

\newcommand{\htco}{\ifmmode {{\rm H}^{13}{\rm CO}^{+} } \else {${\rm H}^{13}
{\rm CO}^{+}$ }\fi}
\newcommand{\hco}{\ifmmode {{\rm H}^{12}{\rm CO}^{+} } \else {${\rm H}^{12}
{\rm CO}^{+}$ }\fi}
\newcommand{\juz}{\ifmmode {{\rm J}=1\rightarrow\0} \else
{J=1$\rightarrow$0}\fi}
\newcommand{\jdu}{\ifmmode {{\rm J}=2\rightarrow\1} \else
{J=2$\rightarrow$1}\fi}
\newcommand{\jtd}{\ifmmode {{\rm J}=3\!\rightarrow\!2} \else
{${\rm J}=3\!\rightarrow\!2$} \fi}
\newcommand{\jcq}{\ifmmode {{\rm J}=5\!\rightarrow\!4} \else
{${\rm J}=5\!\rightarrow\!4$} \fi}
\newcommand{\as}{\ifmmode {^{\scriptscriptstyle\prime\prime}}
        \else $^{\scriptscriptstyle\prime\prime}$\fi}
\newcommand{\am}{\ifmmode {^{\scriptscriptstyle\prime}}
        \else $^{\scriptscriptstyle\prime}$\fi}
\begin{document}
%
\title{Deuterated molecules in DM Tau: DCO$^+$, but no HDO}
\author{St\'ephane Guilloteau \inst{1} \and Vincent Pi\'etu \inst{1}  \and Anne Dutrey \inst{1} \and Michel Gu\'elin
\inst{2}}
\offprints{S.Guilloteau, \email{guilloteau@obs.u-bordeaux1.fr}}
\institute{L3AB, Observatoire de Bordeaux, 2 rue de l'Observatoire, BP 89, F-33270 Floirac, France
 \and IRAM, 300 rue de la piscine, F-38406 Saint Martin d'H\`eres Cedex, France}
\date{Received  /  Accepted }
\authorrunning{Guilloteau, Pi\'etu, Dutrey, \& Gu\'elin}
\titlerunning{No HDO in DM Tau}

\abstract{We report the detection of the J=2-1 line of DCO$^+$ in the proto-planetary disk of  DM~Tau  and
re-analyze the spectrum covering the 465 GHz transition of HDO in this source, recently published by
\cite{Ceccarelli_etal2005}. A modelling of the DCO$^+$ line profile with the source parameters derived from
high resolution HCO$^+$ observations yields a DCO$^+$/HCO$^+$ abundance ratio of $~\simeq 4\times 10^{-3}$, an
order of magnitude smaller than that derived in the low mass cores. The re-analysis of the 465 GHz spectrum,
using the proper continuum flux (0.5 Jy) and source systemic velocity ($6.05 \kms$), makes it clear that the
absorption features attributed to HDO and C$_6$H are almost certainly unrelated to these species. We show that
the line-to-continuum ratio of an absorption line in front of a Keplerian disk can hardly exceed the ratio of
the turbulent velocity to the projected rotation velocity at the disk edge, unless the line is optically very
thick ($\tau >10^4$). This ratio is typically 0.1-0.3 in proto-planetary disks and is $\simeq 0.15$ in DM Tau,
much smaller than that for the alleged absorption features. We also show that the detection of H$_2$D$^+$ in DM
Tau, previously reported by these authors, is only a 2-sigma detection when the proper velocity is adopted. So
far, DCO$^+$ is thus the only deuterated molecule clearly detected in proto-planetary disks. }

\maketitle{}



\section{Introduction}

Deuterated molecules are powerful probes of the physical and chemical state of star forming regions. They are
routinely observed in low mass pre-stellar cores where they are found to be particularly abundant \citep[see
\eg{}][]{Butner_etal1995}. Since the detection of several H-bearing molecules in the protoplanetary disks
surrounding DM~Tau and GG~Tau \citep{Dutrey_etal1997}, the question rose of whether deuterated molecules could
also be observed in proto-planetary disks. \cite{Dishoeck_etal2003} have reported the detection of DCO$^+$ in
the disk of TW Hya. Recently, \cite{Ceccarelli_etal2004,Ceccarelli_etal2005} (hereafter CD04,CD05) have
reported the detection in DM~Tau of H$_2$D$^+$, a key ion in deuterium enhancement reactions, and of deuterated
water, HDO. In this letter, we present the detection of DCO$^+$ in the disk of DM~Tau.

The HDO line reported by CD05 and two lines of C$_6$H present in the same spectrum appear in absorption against
the thermal emission of the dusty disk. These absorptions, if real, forces us to revise our views on the disk
structure, since water is expected to freeze onto grains at the low temperatures ($<25$~K) encountered in such
disks. They also imply very large molecular column densities. We argue in this letter that the low
signal-to-noise ratio of the observations and an overestimation of the continuum flux by a factor 4 cast doubt
on the reality of the HDO and C$_6$H detections.

\section{Observational Data}
\label{sec:obs}

\begin{figure}[t]
\begin{center}
\resizebox{8.0cm}{!}{\includegraphics[angle=0]{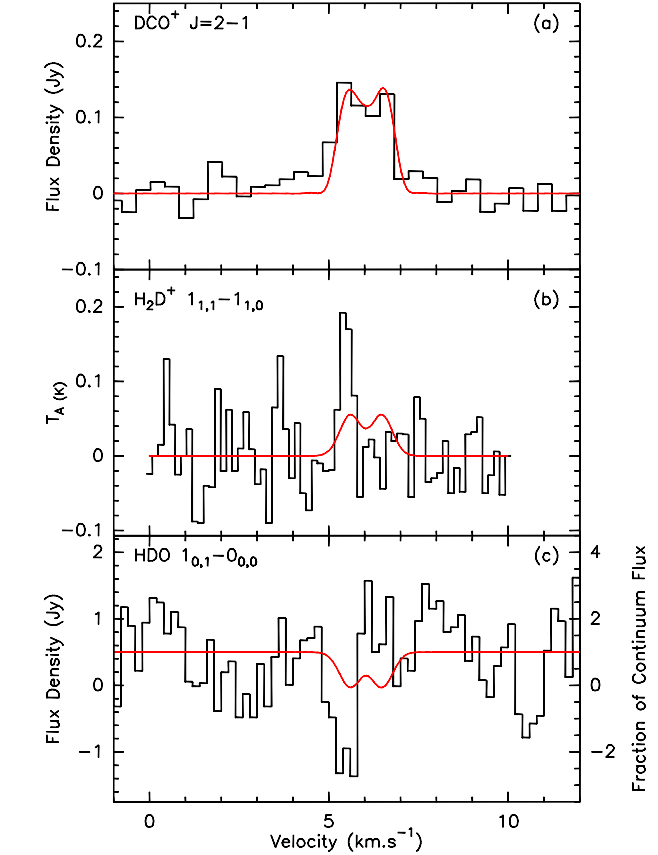}} \caption[Cinematique]{\label{fig:observed} (a)
(c) DCO$^+$ J=2-1 spectrum of DM Tau (this work), superimposed on the best fit double-peaked profile. (b)
H$_2$D$^+$ \unun\ spectrum of DM Tau from CD04, and best fit double-peaked profile. (c) HDO spectrum of DM Tau
from CD05, rescaled with the proper continuum value, superimposed on the best fit double-peaked profile (for
illustration only, see Sec.3). All fit use the same parameters for velocity and line width.}
\end{center}
\end{figure}

We have observed the DCO$^+$ J=2-1 transition at 144.0773 GHz using the IRAM 30-m telescope in Oct 199, Feb and
Sep 2000, using the observing strategy of \citet{Dutrey_etal1997}. The system temperature was about 200 K, and
the total integration time 39 hours. The resulting spectrum is presented in Fig.\ref{fig:observed}a, with the
best fit profile derived using a double gaussian with parameters ($V_\mathrm{LSR} = 5.60$ and $6.48~\kms$ and
line width $\mathrm{FWHM} = 0.7~\kms$) taken from the fit of the \tco~\jdu~ transition given by
\cite{Guilloteau_Dutrey_1994}. The integrated line flux is $0.29 \pm 0.03~\mathrm{Jy}\,\kms$.

CD05 report the detection of the HDO \unzero~ line with a peak line to continuum ratio of about 0.9, a
linewidth of $0.63 \pm 0.18 \kms$, and an LSR velocity of $5.5 \kms$ \citep[using the rest frequency of the HDO
\unzero\ transition, 464.9425 GHz from][]{DeLucia_etal1971}. However, the systemic velocity of the DM Tau disk
is $6.05 \pm 0.02 \kms$ \citep{Guilloteau_Dutrey1998}, and all molecular lines  detected by
\citet{Dutrey_etal1997} exhibit a flat-topped or double-peaked profile typical of Keplerian disks, with a line
width of $\simeq 1.5 \kms$.

Furthermore, CD05 used a continuum flux of 2.0 Jy for DM Tau at 464 GHz (650 $\mu$m), based on a model of the
source presented in their Figure 2. However, the model curve indicates a 464 GHz flux of $\approx 2 \times
10^{-12}$~erg\,cm$^{-2}$\,s$^{-1}$, or $\approx 0.5$~Jy only, in agreement with the two measurements near this
frequency displayed on the same Figure. \cite{Dartois_etal2003} quote $S_\nu(\nu) = 110 (\nu/ 230
\mathrm{GHz})^{2.6}$ mJy in the millimeter range, which yields 630 mJy at 464 GHz. This is an upper limit,
since the SED is expected to flatten above 250 GHz because of the $\nu^2$ contribution of the optically thick
core (as can be seen in Figure 2 of CD05). \cite{Andrews_Williams_2005} measured flux densities of $1.08 \pm
0.05$ Jy at 350 $\mu$m and $240 \pm 10$ mJy at 850 $\mu$m, which, interpolated by a single power law, gives 380
mJy (in this case a lower limit to the flux). Thus, the continuum flux of DM Tau at 464 GHz is most likely
about 0.45 Jy,  more than a factor of 4 below what has been assumed. Fig.\ref{fig:observed}c displays the HDO
spectrum with the line to continuum ratio computed assuming a continuum flux of 0.5 Jy and an antenna gain of
35 Jy/K, as appropriate for the JCMT at 460 GHz.

Using the correct continuum flux and systemic velocity, the fitted equivalent width of the HDO line (which is
the integrated area of the line to continuum ratio)  is $W = -1.6 \pm 1.6 \kms$. Similar results $W = -1.2 \pm
1.3 \kms$ are obtained when fixing the line width to $1.5 \kms$, or by using a double-peaked line profile as
above for DCO$^+$.  The absorption feature assigned to HDO is thus not significant.

The inappropriate velocity of 5.5 $\kms$ was used by CD04 for the detection of the ortho-H$_2$D$^+$ \unun\
line. When analyzed with the 6.05 $\kms$ velocity as above, the integrated area of this line is $0.08 \pm
0.04$~K\,$\kms$, i.e. at best a 2 $\sigma$ detection (see Fig.\ref{fig:observed}b).

\section{Absorption Lines in Keplerian Disks}
\label{sec:absorption}

\begin{figure}[!t]
\begin{center}
\resizebox{6.0cm}{!}{\includegraphics[angle=270]{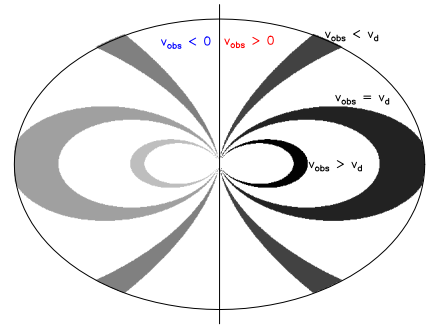}} \caption[Cinematique]{\label{fig:cinematique}
Regions of the disk which yield equal projected velocities $v_\mathrm{obs}$. The outer ellipse is the
projection of the disk outer edge.}
\end{center}
\end{figure}

In proto-planetary Keplerian disks, the line formation process is different from that in an homogeneous static
medium, and this for three reasons: 1) the velocity shear created by the disk rotation 2) the strong surface
density gradient as a function of radius, and 3) the radial temperature gradient.

\begin{figure*}[!th]
\begin{center}
\resizebox{13.0cm}{!}{\includegraphics[angle=270]{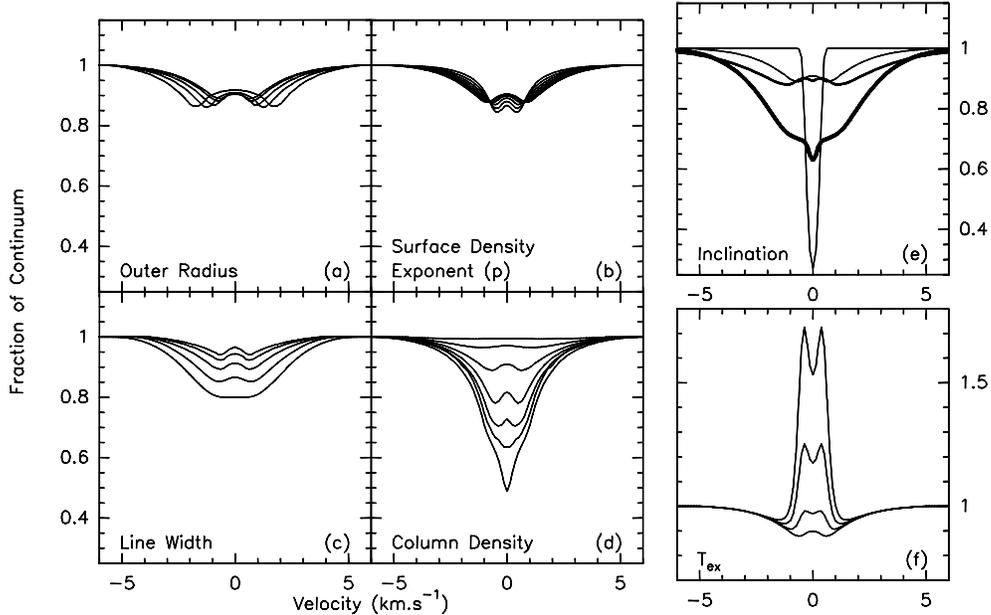}} \caption[model]{\label{fig:model} Predicted
absorption line profile as a function of disk parameters. The intensity is represented in units of the
continuum flux. (a): as a function of outer disk radius ($R_\mathrm{out}$ 50, 100, 200, 400 and 600 AU). (b) as
a function of surface density exponent ($p$ = 0 to 3.0 by 0.5) (c): as a function of intrinsic (local,
turbulent or thermal) line width (0.05, 0.1, 0.2, 0.4 and 0.8 $\kms$). (d): as a function of molecular column
density at 100 AU ($10^{12}$ to $10^{18}$ cm$^{-2}$ by factors of 10). (e): as a function of disk inclination
(from thin to thick: 0, 30, 60 and 90$^\circ$). (f): as a function of excitation temperature (2.7, 3.7, 4.7 and
5.7~K).}
\end{center}
\end{figure*}

The first effect has been discussed in details in the context of line emission from Keplerian disks by
\cite{Horne_Marsh_1986}, for application to cataclysmic binaries. For absorption, the  maximum line to
continuum ratio is limited by the velocity coherence along the line of sight. Unless the disk is seen face on,
only a fraction of the disk projects at any given velocity. This fraction can be estimated with simple
reasoning. From a point at $r,\theta$ (in cylindrical coordinates, with $\theta =0$ at the intersection of the
disk plane with the plane of the sky) in a geometrically thin Keplerian disk, the projected velocity along the
line of sight is
\begin{equation}
V_\mathrm{obs}(r,\theta) = V_\mathrm{LSR} + \sqrt{GM_{*}/r} \sin{i}\cos{\theta}
\end{equation}
We can set $V_\mathrm{LSR} = 0$ by a simple change of referential.  The projection of the rotation velocities
along the line of sight breaks the azimuthal symmetry, but preserves
\begin{eqnarray}
V_\mathrm{obs}(r,\pi -\theta) =   - V_\mathrm{obs}(r,\theta) & \mathrm{and} & V_\mathrm{obs}(r, -\theta) =
V_\mathrm{obs}(r,\theta) \label{eq:vobs}
\end{eqnarray}
allowing to study only the interval $0 < \theta < \pi/2$. The locii of isovelocity is given by
\begin{eqnarray}
r(\theta) &=& ( GM_{*}/V_\mathrm{obs}^2 ) \sin^2{i}\cos^2{\theta}
\end{eqnarray}
Eq.\ref{eq:vobs} implies that the spectrum is symmetric. With a finite local line width $\Delta v$, the line at
a given velocity $V_\mathrm{obs}$ originates from a region included between $r_i(\theta)$ and $r_s(\theta)$ :
\begin{eqnarray}
r_i(\theta) &=& \frac{GM_{*}}{(V_\mathrm{obs}+\Delta v/2)^2}\sin^2{i}\cos^2{\theta} \\
r_s(\theta) &=& \min\left[R_\mathrm{out},\frac{GM_{*}}{(V_\mathrm{obs}-\Delta v/2)^2}
\sin^2{i}\cos^2{\theta}\right]
\end{eqnarray}
assuming a rectangular line shape for simplification. We define $v_d$ as :
\begin{eqnarray}
v_d &=&  \sqrt{GM_{*} / R_\mathrm{out}} \sin{i} \label{eq:vd}
\end{eqnarray}
where $R_\mathrm{out}$ is the disk outer radius. In general, unless $i$ is small, $\Delta_v\ll v_d$. Then for $
V_\mathrm{obs} < v_d+\Delta v/2$ , $r_s(\theta = 0) = R_{out}$. For larger velocities, $ V_\mathrm{obs} > v_d +
\Delta v/2 $, $r_s$ is smaller and given by
\begin{eqnarray}
r_s(\theta = 0) &=& GM_{*} \sin^2{i} / (V_\mathrm{obs}-\Delta v/2)^2 \label{rs}
\end{eqnarray}
Figure \ref{fig:cinematique} indicates the regions of equal projected velocities for 6 different values:
$V_\mathrm{obs}>v_d$, $V_\mathrm{obs} = v_d$, $V_\mathrm{obs}<v_d$, and their symmetric counterpart at negative
velocities. Fig.\ref{fig:cinematique} shows that the fraction of the disk covered by the gas at velocities
$V_\mathrm{obs}<v_d$ is of order $\Delta v / v_d$, and drops very rapidly for larger velocities. Accordingly,
the line to continuum ratio of absorption lines cannot exceed this value, and will be smaller for optically
thin lines.

For a more quantitative evaluation, we used the radiative transfer code developed by \cite{Dutrey_etal1994} to
simulate the expected line shape for an absorption line from a Keplerian disk with parameters (inclination,
dust emission, $\Delta v$, scale height) similar to those of DM Tau. To perform this simulation, we assumed
that the HDO abundance is constant, and that the surface density is a power law as function of radius
$\Sigma(r) = \Sigma_{100} (r/100~\mathrm{AU})^{-p}$ with exponent $p=1.5$. We did not assume LTE, but because
HDO is difficult to excite by collisions, we assumed instead that the levels are populated only by the
cosmological background, with a line excitation temperature of 2.7 K. The continuum emission comes from dust
with a power law distribution for the temperature
 $T(r) = 20 (r / 100 \mathrm{AU})^{-0.3}$~K,
with the same surface density power law. Unless otherwise specified, the disk inclination is set to $30^\circ$,
the local line width to 0.2 $\kms$, the outer radius to $600$~AU, the orbital velocity at 100~AU to 2.10
$\kms$, corresponding to a stellar mass of $0.50 \msun$, and the HDO column density at 100 AU to
$10^{14}$~cm$^{-2}$.

Fig \ref{fig:model}a-f show the expected profile for different disk parameters. Since the continuum brightness
is highly centrally peaked, the line to continuum ratio depends significantly neither on the disk outer radius,
nor on the surface density exponent (Fig \ref{fig:model}a-b). Only the separation between the two peaks depends
on these parameters. Fig \ref{fig:model}e shows that the line shape depends strongly on the disk inclination.
For a face on disk, the line is single-peaked, and its width depends on the local line width (with a
$\sqrt{\ln(\tau)}$ opacity broadening factor). For moderate inclination, the line becomes double-peaked, with
the two peaks corresponding to the projected rotation velocity near the disk outer edge. For an edge-on disk,
the opacity towards the central star (thus at a zero projected velocity) becomes important, and the central dip
is filled again when the column density is high enough (as in this example).

\begin{figure}[!t]
\begin{center}
\resizebox{8.0cm}{!}{\includegraphics[angle=270]{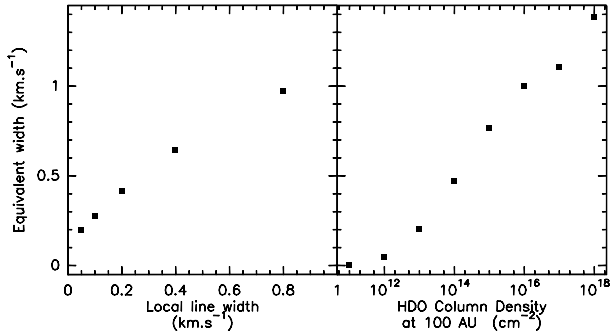}} \caption[Width]{\label{fig:width} Left:
Equivalent line width of the absorption line as function turbulent line width for a column density of $10
\times 10^{14}$ cm$^{-2}$. Right: HDO column density for a turbulent line width of 0.2 $\kms$. Note the
logarithmic scale for the column density.}
\end{center}
\end{figure}

The two most significant effects are due to the intrinsic line-width (Fig.\ref{fig:model}c) and the molecular
column density (Fig.\ref{fig:model}d). The equivalent width  increases nearly linearly with the local line
width, but only with the logarithm of the column density (through the opacity broadening effect), as can be
seen in Fig.{\ref{fig:width}}.

We used in Fig.\ref{fig:model}a-e a minimal excitation temperature, 2.7~K.  For increasing values of
$T_\mathrm{ex}$, due e.g. to collisions, the absorption at zero velocity will first diminish, then reverse to
emission, the line wing reversing to emission at slightly higher values (because the wings originate from the
inner disks, where the dust brightness temperature is higher), as shown in Fig.\ref{fig:model}f. Non uniform
mixing with the dust will affect the line intensity, although in general weakly, see e.g.
\cite{Walterbos_1988}, but will not significantly affect the line shape which is dictated by the velocity
structure.

Finally, we note that the simulated profiles could also be applied to other molecular transitions, provided the
column density is changed to the appropriate scale: here, a column density of $10^{14}$ cm$^{-2}$ corresponds
to a typical opacity of 4 at 100 AU.

\section{Discussion}

We have shown in Sec.\ref{sec:obs} that there is little evidence that the absorption feature identified by CD05
with the $1_{01}-0_{0,0}$ line of HDO (see Fig.\ref{fig:observed}c) is real and arises in the disk of DM Tau,
and we have set  in Sec.\ref{sec:absorption} further constraints on the expected line profiles in such a disk.
Two additional absorption features at 464917.2 MHz and 465051.1 MHz, observed in the same spectrum, were
identified by CD05 with lines of C$_6$H. This identification also is almost certainly incorrect. Although two
C$_6$H transitions are listed at these frequencies in the JPL Spectral line catalog \citep{JPL_catalog}, these
are cross-ladder transitions that connect the $^2\Pi_{1/2}$ upper spin state of C$_6$H to the $^2\Pi_{3/2}$
ground spin state. These transitions are intrinsically weak and originate from levels with high energies
($E/k>80$ K): they are very unlikely to be seen in DM Tau, especially in absorption. Mostly, their frequencies
depend on the spin orbit constant $A$, whose value has not been directly measured, and are uncertain by several
hundred MHz \citep{Pearson_etal1988}, which prevents any sensible identification.

The detection of H$_2$D$^+$ in DM Tau reported by CD04 is also not secure, as it is only at the $2 \sigma$
level when fitted with an appropriate line profile. Since the tentative detection in TW Hya reported in the
same paper has also not been confirmed, but remains consistent with a 2 $\sigma$ upper limit of 0.3 K\,$\kms$
from the JCMT (Thi \& van Dishoeck 2004, cited by CD04), this re-analysis casts doubt on the detection of
H$_2$D$^+$ in proto-planetary disks at large.

This brings the number of deuterated species firmly detected in proto-planetary disks down to only
one, DCO$^+$, in two sources: in TW Hya, through the detection of the DCO$^+$ J=5-4 transition by
\cite{Dishoeck_etal2003}, and in DM Tau (this work).


We have analyzed the DCO$^+$ line using the DM Tau disk parameters derived from high angular resolution
observations of the HCO$^+$ J=1-0 line performed with the IRAM interferometer \citep{Pietu_etal2006}. We find
that this line profile corresponds to a DCO$^+/$HCO$^+$ ratio of $4.0 \pm 0.9 \times 10^{-3}$, provided this
ratio is constant throughout the disk.  Note that a derivation which does not take into account the specificity
of the line formation process in a disk (e.g. which would assume HCO$^+$ and DCO$^+$ lines to be optically
thin, or would not take into account the density and temperature gradients) would give a completely different
DCO$^+$/HCO$^+$ ratio. When no resolved image is available, the analysis must still use an approximate disk
structure, as done for example by \cite{Dutrey_etal1997}. The DCO$^+$/HCO$^+$ abundance ratio derived for DM
Tau is smaller than in typical dark clouds or (cold) proto-stellar envelopes \citep[$0.01 - 0.07$, see][]
{Butner_etal1995}.

Finally, Sec.\ref{sec:absorption} shows that the detection of absorption lines from proto-planetary disks
require very high sensitivities, and that such lines are very insensitive tracers of the total column density.
At 464 GHz, with 1 $\kms$ velocity resolution, and a system temperature of 400 K, a good sub-mm 12-m antenna
such as APEX would reach a $1 \sigma$ noise of 0.3 Jy in 1 hour. This is 3 times more than the maximum
absorption signal that could be expected in typical proto-planetary disks (line to continuum ratio 0.2, flux
density 0.5 Jy). A 4 $\sigma$ detection would require more than 100 hours. Only ALMA, with its 50 antennas,
would reach enough sensitivity, $\simeq 7$~mJy in 1 hour for its minimal angular resolution. However, this
resolution is already 0.7$''$ at this frequency, so that the continuum source would be partially resolved.

 \acknowledgements{We
thank Pierre Hily-Blant for helping with the data reduction of the 30-m observations.}
\bibliography{hk032}
\bibliographystyle{aa}
\end{document}